\documentstyle[art12]{article}

\def\mos{microorganism}
\def\pa{\partial}

\begin{document}

\pagestyle{empty}

\noindent
ICHEP 94 Ref. gls0084 \hfill OCHA-PP-43

\noindent
Submitted to Pa 19 \hfill NDA-FP-15

\noindent
\ \hskip 2.4cm Pl 11 \hfill June (1994)

\vfill

 \begin{center}
     {\LARGE  Swimming of Microorganism \\
              and the String- and Membrane- like Algebra}\\
      \vfill
    {\large Masako KAWAMURA,  Akio SUGAMOTO}\\
    \vspace{0.1in}
        {\it Department of Physics, Faculty of Science\\
         Ochanomizu University \\
        1-1 Otsuka 2, Bunkyo-ku, Tokyo 112, Japan}\\
        \vspace{0.2in}
        and \\
        \vspace{0.2in}
        {\large  Shin'ichi NOJIRI} \\
        \vspace{0.1in}
        {\it Department of Mathematics and Physics \\
        National Defence Academy \\
        Yokosuka, 239, Japan}
 \end{center}
 \vfill
\begin{abstract}
Swimming of microorganisms is further developed from a viewpoint of strings
and membranes swimming in the incompressible fluid of low Reynolds number.
In our previous paper the flagellated motion was analyzed in two dimensional
fluid, by using the method developed in the ciliated motion with the
Joukowski transformation.
This method is further refined by incorporating the inertia term of fluid
as the perturbation. Understanding of the algebra controlling the
deformation of microorganisms in the fluid is further developed, obtaining
the central extension of the algebra with the help of the recent progress
on the $W_{1+\infty}$ algebra. Our previous suggestion on the usefulness
of the $N$-point string- and membrane-like amplitudes for studying the
collective swimming motion of $N-1$ microorganisms is also examined.
\end{abstract}

\newpage
\pagestyle{plain}
\setcounter{page}{1}
\section*{Introduction}

In our previous paper\cite{1} we have studied swimming of microorganisms
viewed from string and membrane theories.
Our target is still the understanding of the following ways of microorganisms'
swimming \cite{2}:There exist only three different universality classes of
the swimming ways of microorganisms; (1) Swimming with cilia is adopted by
the spherical organisms with the length scale of
$20 \sim 2 \times 10^4 \mu$m, an example of which is {\it paramecium}; (2)
the smaller microorganisms with the size of $1 \sim 50 \mu$m swim with
flagella, an example of which is the {\it sperm}; (3) the bacteria with
the size of $0.2 \sim 5 \mu$m swim with bacterial flagella the motion of
which resembles the screwing of the wine-opener. Why is it possible that such
simple classification be realized in the swimming problem of microorganisms?

Recapitulating our previous work\cite{1}, we will comment on a few newly
 obtained results. Starting from the work by Shapere and Wilczek\cite{3},
 we have developed the ciliated and flagellated motions and studied the
 microorganisms' swimming from the viewpoint of algebraic structure existing
in the deformation operation of the microorganisms in the low Reynolds liquid,
as well as from the swimming dynamics of a group of microorganisms, which may
be connected to the $N$-point correlation function of string and membrane
theories.

The microorganisms with the length scale $L \ll 1$, the Reynolds number $R$
satisfies $R \ll 1$, so that the hydrodynamics in this case leads to the
following equations of motion for the incompressible fluid:
\begin{equation}
  \nabla \cdot  {\bf v} = 0,    \label{2}
\end{equation}
and
\begin{equation}
  \Delta {\bf v}  = \frac{1}{\mu} \nabla p,      \label{3.1}
\end{equation}
or equivalently
\begin{equation}
  \Delta (\nabla \times {\bf v}) = 0,        \label{3.2}
\end{equation}
where $p$ is the pressure and ${\bf v}(x)$ is the velocity field of the
fluid. The surface of a microorganism swimming in the $D=3$ dimensional
fluid of the real world forms a closed membrane at a fixed times $t$, the
position of which can be parametrized by introducing $(D-1)$ parameters
$\xi^i$ $(i=1, \cdots, D-1)$ as
$X^{\mu} = X^{\mu}(t; \xi^1, \cdots, \xi^{D-1})$. It is sometimes
instructive to consider the $D=2$ dimensional fluid. Then, the surface
of a ciliate (flagellate) becomes a closed (open) string and its position
can be described by a complex number
\[  Z = x^1 + ix^2 = Z(t; \theta),  \]
with $-\pi \leq \theta \leq \pi$. In the sticky fluid of $\mu\neq 0$,
there is no slipping between the surface of a microorganisim  and the
fluid, namely, we have the matching condition
\begin{equation}
 {\bf v}({\bf x}={\bf X}(t; \xi))= \dot{{\bf X}}(t; \xi),    \label{4.1}
\end{equation}
or in the general coordinate system with the metric tensor $g^{\mu \nu}(x)$
\begin{equation}
   g^{\mu \nu}(x)v_{\mu}(x)|_{x = X(t; \xi)} = \dot{X}^{\nu}(t; \xi).
   \label{4.2}
\end{equation}

\section{The ciliated motion in the two dimension fluid}

The ciliated motion can be viewed as a small but time-dependent
deformation of a unit circle in a properly chosen scale,
\begin{equation}
 Z(t, \theta) = s + \alpha(t, s),       \label{5}
\end{equation}
where $s=e^{i\theta}$ and $\alpha(t, s)$ is arbitrary temporally
periodic function with period $T$ satisfying $|\alpha(t, s)| \ll 1$
with $-\pi \leq \theta \leq \pi$. The complex representation of the
velocity vector $v_\mu$ can be denoted as
\begin{eqnarray}
2v_{\bar{z}}(z, \bar{z})& = &(v_1 +iv_2)(z, \bar{z})  \label{5.1}    \\
        2v_z(z, \bar{z})& = &(v_1 - iv_2)(z, \bar{z}).  \label{5.2}
\end{eqnarray}
By estimating the translational and rotational flows at spacial infinity
caused by the deformation of the cilia, we have obtained $O(\alpha^2)$
expression of the net translationally swimming velocity
$v_{T}^{(\rm cilia)}$ of the ciliated microorganism as follows:
\begin{eqnarray}
    \lefteqn{ 2v_{T}^{({\rm cilia})}  =  -\dot{\alpha}_0 (t)} \nonumber   \\
      &  & + \sum_{n \leq 1} n (\dot{\alpha}_n \alpha_{-n+1}
      - \overline{\dot{\alpha}_n} \alpha_{n-1} - \overline{\dot{\alpha}_n}
      \overline{\alpha_{-n+3}})         - \sum_{n>1} n \dot{\alpha _n}
      \overline{\alpha _{n-1}},         \label{18}
\end{eqnarray}
where $\alpha_n (t)$ is defined by $\alpha (t, s) =
\sum_{n=-\infty}^{+\infty} \alpha_n (t) s^n$.
On the other hand, the net angular momentum $v_{R}^{({\rm cilia})}$
gained by the microorganism from the fluid becomes
\begin{eqnarray}
 \lefteqn{ 2v_{R}^{({\rm cilia})}  = -{\rm Im} \left\{ \dot{\alpha}_1 (t)
 \right.}   \nonumber      \\
    &  & - \left. \sum_{n \leq 1} n (\dot{\alpha}_n \alpha_{-n+2}
    - \overline{\dot{\alpha}_n} \alpha_{n} - \overline{\dot{\alpha}_n}
    \overline{\alpha_{-n+2}})        +  \sum_{n>1} n \dot{\alpha _n}
    \overline{\alpha _n} \right\}.                       \label{19}
\end{eqnarray}
The net translation and rotation resulted after the period $T$ come from
$O(\alpha^2)$ terms since the $O(\alpha)$ terms cancel after the time
integration over the period.

\section{The flagellated motion in two dimensional fluid}

Microorganisms swimming using a single flagellum can be viewed as an
open string with two endpoints, H and T, where H and T represent the
head and the tail-end of a flagellum, respectively. Our discussion
will be given by assuming that the distance between H and T is
time-independent and is chosen to be 4 in a proper length scale.
This assumption can be shown to be valid for the flagellated motion
by small deformations in the incompressible fluid.
Then, at any time $t$, we can take a complex plane of $z$,
where H and T are fixed on $z=2$ and $-2$, respectively. This coordinate
system $z$ can be viewed as that of the space of {\it standard shapes} of
Shapere and Wilczek. Time dependent, but small deformation of the
flagellate can be parametrized as
\begin{equation}
   Z(t, \theta) = 2(\cos \theta + i\sin \theta \alpha (t, \theta)) ,
   \label{20}
\end{equation}
where the small deformation $\alpha (t, \theta)$ can be taken to be a
real number\footnote{When $\alpha$ is taken to be a complex number,
the length of the
flagellum is locally changeable at $O(\alpha)$.
For such an elastic flagellum, we have similar results to that of the
ciliated motion. In case of real $\alpha$, its length is locally
preserved at $O(\alpha)$, giving a non-elastic flagellum, which is
the more realistic one.}
satisfying
\begin{equation}
  \alpha (t, \theta) = -\alpha (t, -\theta).        \label{21}
\end{equation}
Here, we parametrize the position of the flagellum twice, starting from the
endpoint T at $\theta = -\pi$, coming to the head H at $\theta = 0$, and
returning to T again at $\theta = \pi$.
Motion of the two branches corresponding to $-\pi \leq \theta \leq 0$
and $\pi \geq \theta \geq 0$ should move coincidentally, which requires
the condition (\ref{21}).
The Joukowski transformation $z = z(w)= w + w^{-1}$, separates the two
coincident branches in the $z$ plane to form lower and upper parts of a
unit circle in the $w$ plane, outside domain of which we are able to
study the swimming problem of the flagellate in a quite similar
fashion to that of the ciliate.
The parametrization of our microorganism in the $w$ plane
corresponding to Eq.(\ref{20}) is now
\begin{equation}
  W(t, \theta) = e^{i \theta}(1 + \alpha (t, \theta)) + O(\alpha^2).
       \label{22}
\end{equation}
Using the mode expansion satisfying Eq.(\ref{21}),
\begin{equation}
  \alpha (t, \theta) = \sum_{n=1}^{\infty} \alpha_n (t) \sin n\theta,
  \label{31}
\end{equation}
we are able to determine the net swimming velocity
${v_T}^{\rm(flagella)}$ gained by the flagellate motion of
microorganisms:
 \begin{equation}
  2v_T ^{(\rm flagella)} = -i \dot{\alpha}_1
  - \sum_{m \geq 1} m \alpha_m \dot{\alpha}_{m+1} +
  \sum_{m \geq 2} m \alpha_m \dot{\alpha}_{m-1} ,   \label{34}
\end{equation}
On the other hand, the angular momentum ${v_R}^{(\rm flagella)}$ is given by
\begin{equation}
   2{v_R}^{(\rm flagella)}  = - \frac{1}{2} \dot{\alpha}_2.   \label{35}
\end{equation}
After the time integration over the period $T$,
${v_R}^{(\rm flagella)}$ vanishes since in our first order
approximation, the length of the flagellum is fixed in the
incompressible fluid. Therefore the second order approximation
is necessary for the non-vanishing ${v_R}^{(\rm flagella)}$.

\section{The selection rules and the symmetry of microorganisms' swimming}

Even though the results in Eqs.(\ref{18}), (\ref{19}),
(\ref{34}) and (\ref{35}) are obtained perturbatively, we are able to
read from them the characteristics of the \mos s' swimming; In order
for the ciliates to swim or rotate, they need the coexistence of the
two different Fourier modes of $n_1$, $n_2$. The selection rules for
the allowed ($n_1, n_2$) conbinations are
\begin{eqnarray}
  i)& \hspace{0.2cm}|n_1 + n_2|=3 \hspace{0.3cm} {\rm or} \hspace{0.3cm}
  |n_1 \pm n_2| =1 \hspace{0.3cm}& {\rm for\ the\ cililated\ translation}
  \label{36.1}
  \\
  ii) & \hspace{0.3cm} n_1=n_2 \hspace{0.3cm} {\rm or} \hspace{0.3cm}
  |n_1+ n_2|=2    \hspace{1cm} & {\rm for\ the\ ciliated\ rotation},
   \label{36.2}
\end{eqnarray}
The corresponding selection rules for the flagellate motion are
\begin{equation}
   iii) \hspace{1cm} |n_1 - n_2| =1  \hspace{3cm}
  {\rm for\ the\ flagellate\ translation},
  \label{37.1}
\end{equation}
where the Fourier modes are $\sin n\theta$ in this case.

Viewing these selection rules, we are tempted to elucidate the
algebraic structure possibly existing in the background of the
swimming mechanism. It is similar to the Virasoro algebra, but
is different from it. For such a purpose, introduction of the
{\lq\lq}action" will be convenient. The {\lq\lq}action" $S$
reproducing the classical equations of motion of the swimming of
$N$ \mos s in the incompressible liquid with low Reynolds number
may be given by
\begin{eqnarray}
  S_N & = & \sum_{i=1}^{N} \int dt\, \int d^{D-1}\xi_{(i)}\, P_{\mu}^{(i)}
  (t; \xi_{(i)}) \left[ \dot{X}^{\mu}_{(i)}(t; \xi_{(i)})
  - v^{\mu}(X_{(i)}(t; \xi_{(i)})) \right]    \nonumber     \\
    &   & \mbox{} +\frac{1}{2 \pi \alpha'} \int d^D x\, \sqrt{g(x)}\,
    \left[ -\frac{1}{\mu}p(x) \partial_{\nu} v^{\nu}(x) + \frac{1}{4}
    \omega_{\mu \nu}(x) \omega^{\mu \nu}(x) \right]     \label{38}
\end{eqnarray}
where the velocity field $\omega_{\mu \nu}(x)$ is given by
\begin{equation}
   \omega_{\mu \nu}(x) = \partial_{\mu} v_{\nu} - \partial_{\nu}v_{\mu},
              \label{39}
\end{equation}
whose $D=2$ expression is $\omega_{z, \bar{z}}(z, \bar{z})
= \partial_z v_{\bar{z}} - \partial_{\bar{z}}v_z$.

We have introduced the parameter $\alpha'$ so as to make $S$
dimensionless, where $\alpha'$ has the dimension of
$({\rm Length})^D({\rm Time})^{-2}$. The reason why we have used
the notation $\alpha'$, familiar in the string theories to
describe the Regge slope, will be understood later. The Lagrange
multiplier fields of $P_{\mu}^{(i)}(t; \xi_{(i)})$ $(i=1, \cdots, N$,
$\mu=1, \cdots, D)$ guarantee the matching condition of (\ref{4.1})
or (\ref{4.2}) for $i$-th \mos\ at any time, and the pressure $p(x)$
is also such multiplier giving the incompressibility given in
Eq.(\ref{2}). The field equation (\ref{3.1}) can be easily
reproduced. In the action (\ref{38}), time $t$ appears only in the
first term of representing the matching conditions, that is, the
time evolution is triggerd only by the self-motion of the \mos s,
of which influence spreads instantaneously over the whole space and
causes the change of the fluid velocity there. Because of the lacking
of the kinetic term, we may call $S$ as the
 {\lq\lq}action". The additional metric contribution such as
 $\sqrt{g(x)}$ is only relevant for the curved space, an example of
 which has appeared in the flagellate swimming on the $w$ plane.
 The later discussion is given for the flat metric.

Now, we will define the following local transformation at a fixed time $t$:
\begin{eqnarray}
    \delta \dot{X}^{\mu}_{(i)} & = & \lambda^{\mu}
    (X_{(i)} (t; \xi_{(i)}))                \label{40.1}      \\
    \delta P_{\mu}^{(i)} & = & 0           \label{40.2}    \\
    \delta v^{\mu} (x) & = & \lambda^{\mu}(x)       \label{40.3}    \\
    \delta p(x) & = & \kappa (x),          \label{40.4}
\end{eqnarray}
where we have assumed that the transformation parameters
$\lambda^{\mu}(x)$ and $\kappa (x)$ are restricted by
the equations of motion,
\begin{equation}
  \partial_{\mu} \lambda^{\mu}(x) = 0,  \hspace{0.2in} {\rm and}
  \hspace{0.2in}  \partial_{\mu}{\lambda^{\mu}}_{\nu}(x)= \frac{1}{\mu}
  \partial_{\nu} \kappa(x),   \label{41}
\end{equation}
where $\lambda_{\mu \nu} \equiv \partial_{\mu}\lambda_{\nu}
- \partial_{\nu}\lambda_{\mu}$ is the vorticity for $\lambda^{\mu}$.
Meaning of the transformations (\ref{40.1})$\sim$(\ref{40.4}) are
quite simple; the deformation of the \mos s (\ref{40.1}) triggers
the increase of the velocity field (\ref{40.3}) and of pressure
(\ref{40.4}) so that they can be consistent with the incompressible
fluid dynamics of the low Reynolds number. It is also important to
note that the succession of these time-independent transformations
result in the time evolution of our problem. Therefore, the
transformations (\ref{40.1})$\sim$(\ref{40.4}) resemble the ordinary
canonical transformation generated by the Hamiltonian.

The transformation also generates the deformation of the shapes
of \mos\ (\ref{40.1}). The generator of this deformation per unit
time can be written as    \begin{equation}
  \hat{L}_{[\lambda^{\nu}]} \equiv \int d^{D-1} \xi \, \lambda^{\mu}
  (X(t; \xi))\frac{\delta}{\delta X^{\mu}(t, \xi)},
  \label{50}
\end{equation}
which gives  \underline{the volume (area for $D=2$) preserving
diffeomorphisms} owing solely to the incompressibility condition
in Eq.(\ref{41}).
  The second condition in Eq.(\ref{41}) adds the further restriction
  on $\hat{L}_{[\lambda^{\nu}]}$: By the help of the stream function
  $\sigma_{\lambda}(x)$,
   the incompressibility  condition is automatically satisfied
   through $\lambda^{\mu} = \epsilon^{\mu \nu \lambda}
   \partial_{\nu} \sigma_{\lambda}$ for $D=3$,
   so that the Eq.(\ref{41}) becomes the constraint on the stream function
\begin{equation}
  \Delta (\Delta g_{\mu \nu} - \partial_{\mu}\partial_{\nu})
  \sigma^{\nu}(x)=0,    \label{51.1}
\end{equation}
or in $D=2$, in terms of the only non-vanishing component $\sigma=\sigma_3$
\begin{equation}
  \Delta^2 \sigma (x)= 0.     \label{51.2}
\end{equation}
This constraint has been already solved generally in $D=3$ and $D=2$ fluid.

The deformation operator of the fluid corresponding to (\ref{50})
\begin{equation}
  L_{[\lambda^{\nu}]} \equiv \lambda^{\mu}(x)\frac{\partial}
  {\partial x^{\mu}},
   \label{57}
\end{equation}
becomes in $D=2$ as
\begin{eqnarray}
  L_{\sigma}& = &2 (\lambda_{\bar{z}} \partial_z
  + \lambda_z \partial_{\bar{z}})    \nonumber   \\
            & = & 2(\partial_{\bar{z}}\sigma \partial_z
            - \partial_z \sigma \partial_{\bar{z}}),
        \label{58}
\end{eqnarray}
where the stream function $\sigma (z, \bar{z})$ contains
$z\bar{z}^k$, $z^k \bar{z}$, $\ln z$, $\ln \bar{z}$,
and $\bar{z}{\ln}z\bar{z}$ terms ($k$ : integer).
Then, $L_{\sigma}$ can be considered as a Liouville operator of a
dynamical system moving in the phase space of $(z, \bar{z})$, having
$-\sigma(z, \bar{z})$ as its Hamiltonian. Invariance of the phase
volume during the temporal evolution of the dynamical system shows
that $L_{\sigma}$ is really the area preserving diffeomorphism. The
commutation relation is simple, namely
\begin{equation}
  [L_{\sigma_1}, \:L_{\sigma_2}] = - L_{ \{ \sigma_1, \: \sigma_2 \} },
    \label{60}
\end{equation}
where the $\{ \sigma_1, \: \sigma_2 \}$ is the Poisson bracket defined by
\begin{equation}
   \{ \sigma_1, \: \sigma_2 \} = \partial_z \sigma_1
   \partial_{\bar{z}} \sigma_2 - \partial_{\bar{z}} \sigma_1
   \partial_z \sigma_2.
   \label{61}
\end{equation}

Recently we have tried to obtain a closed algebra including $\ln z$ and
$\ln \bar{z}$ in $\sigma (z, \bar{z})$.
The algebra so obtained consists of
$L_{(l, m, n)} \equiv L_{z^l \bar{z}^m (\ln z \bar{z})^n}$, and
$M \equiv L_{\ln z - \ln \bar{z}}$; They satisfy
\begin{eqnarray}
[L_{(l, m, n)}, \ L_{(p, q, r)}] & = & -(lq-mp)\,
L_{(l+p-1,\,  m+q-1,\,  n+r)} \nonumber  \\
    & & +(mr+np-lr-nq)\, L_{(l+p-1, \, m+q-1, \, n+r-1)} \nonumber  \\
  & & + c \, (m-l)\,
  \delta _{l+p, \, 0} \, \delta _{m+q, \, 0} \, \delta _{n+r, \, 0}
  \ \  , \\
  \left[ M, \ M\right] & = & 0 \ \ ,
\end{eqnarray}
and
\begin{eqnarray}
[M, \ L_{(l, m, n)}] & = & -(l+m)\, L_{(l-1, m-1, n)}
- 2n \, L_{(l-1, m-1, n-1)}  \nonumber \\
& & \ \ \ \ + \frac{1}{2} c \, \delta_{l, \, 0} \, \delta_{m, \, 0} \,
\delta_{n, \, 0}.
\end{eqnarray}
In the above expression we add the central charge $c$, corresponding to the
possible central extension of the algebra in which the Jacobi identities are
kept to hold and the generators are understood to be properly redefined.

The reason why the $\ln z$ or $\ln \bar{z}$ is permitted in
the stream function $\sigma (z, \bar{z})$ is that the existing
singularities at $z=0$ can be
hidden inside the body of the \mos\ itself.
Therefore, if we are not interested in the circulation flow of the
fluid(topological flow) around the \mos\, we can ignore the logarithmic
contribution in $\sigma$.
In that case, the algebra becomes $w_{1+ \infty}$:
With the notation $T_{n, \, m} = L_{(n, \, m,\, 0)}$, we have\footnote{
For the ordinary $W_{1+\infty}$ algebra, $m\geq 0$ but in our algebra of
microorganisms' swimming, $m$ can be negative.}
\begin{eqnarray}
[T_{n, \, m}, \ T_{k, \, l}] & = & -(nl - mk) \, T_{n+k-1, \, m+l-1}
\nonumber  \\
  & & + c \, (m-n)\, \delta_{m+k,\, 0} \, \delta_{m+l, \, 0}.
  \label{star}
\end{eqnarray}
There is a discrepancy between the definition of the $W_{1+ \infty}$
and the above expression(\ref{star}).
Corresponding to the classical generator $z^k D^n$ of $W_{1+ \infty}$ algebra,
we define the quantum version of the generator as
$V^n _k = W(z^k D^n) \ \displaystyle \left[ D = z
\frac{\partial}{\partial z} \right ]$.
Then, the operator $\tilde{T}_{n , \, m} = \alpha ^{n+m-2} W(z^{n-m}D^m)$
satisfy the algebra (\ref{star}) in the limit of $\alpha \rightarrow 0$ for
$c=0$.

Recently the representation theory of this $W_{1+ \infty}$ algebra is
progressing considerably \cite{4,5,6,7,8,9} so that we are wishing to
apply it to our problem.
Successive application of the infinitesimal deformation on the \mos\
forms a
swimming motion.
On the other hand a representation of the $W_{1+ \infty}$ algebra is obtained
by a successive operation of raising operators
$V_{-r}^n \, (n \geq 1, \, r \geq 1)$ on a highest weight state $| \lambda >$
which is characterized by a set of eigenvalues for
$V_0 ^n \, (n = 1, 2,  3, \ldots)$.
This conceptual correspondence between the swimming motion of \mos s\ and the
representation theory of the algebra, both of which are controlled by the
area-preserving diffeomorphisms, may be useful to understand algebraically
the typical pattern of the \mos s'\  swimming.

\section{Small but Non-Vanishing Reynolds number}

Next, we will study the case in which the Reynolds number $R$ is small
but non-vanishing.
This has not been studied in our previous work \cite{1}.
In fact the length size $L$ of the \mos s\  ranges from $\mu$m to mm.
Therefore, for the larger \mos s\  of $L=O(1\rm mm)$, $R$ of $O(1)$
can not be eliminated from the beginning.
The incompressible fluid with $R \neq 0$ has the following equation of motion:
\begin{equation}
\frac{\rho}{\mu} \left\{ \dot{\bf v}
+ ( {\bf v} \cdot \nabla) \bf v \right\}
= - \frac{1}{\mu} \nabla p + \Delta {\bf v} ,
  \label{kansei}
\end{equation}
or
\begin{equation}
\epsilon \{ \nabla \times \dot{\bf v} + \nabla
\times (({\bf v} \cdot \nabla) {\bf v}) \} = \Delta (\nabla \times {\bf v}),
\label{kansei2}
\end{equation}
with $\epsilon = \rho / \mu$.
This equaiton reduces to eq.(\ref{3.1}) or (\ref{3.2}) for $R=0$.
We will use $\epsilon$ as an expansion parameter of incorporating small
perturbation from the non-vanishing $R$.
Expansion of $\bf v$ in terms of $\epsilon$, ${\bf v} = {\bf v}^{(0)}
+ {\bf v}^{(1)} + \cdots $, leads to
\begin{eqnarray}
   \Delta (\nabla \times {\bf v}^{(0)}) & = & 0  \label{kansei3.1} \\
          \Delta (\nabla \times {\bf v}^{(1)}) & = & \epsilon \{
          \nabla \times \dot{\bf v}^{(0)} + \nabla \times (({\bf v}^{(0)}
          \cdot \nabla) {\bf v}^{0}) \}  \label{kansei3.2}
\end{eqnarray}
for $D=3$, but for $D=2$ they can be written as
\begin{equation}
4\pa_z \pa_{\bar{z}} (\pa_z v_{\bar{z}}^{(0)} - \pa_{\bar{z}} v_z^{(0)}) =0
\label{kansei4}
\end{equation}
\begin{eqnarray}
\lefteqn{4\pa_z \pa_{\bar{z}} (\pa_z v_{\bar{z}}^{(1)}
- \pa_{\bar{z}} v_z^{(1)})}  \nonumber  \\
 & & = \epsilon \left\{
   \begin{array}{l}
    \pa_z \dot{v}_{\bar{z}}^{(0)} - \pa_{\bar{z}}\dot{v}_z^{(0)} \\
    + 2 \pa_z ((v_z^{(0)} \pa_{\bar{z}} + v_{\bar{z}}^{(0)} \pa_z)
    v_{\bar{z}}^{(0)}) - 2 \pa_{\bar{z}} ((v_z^{(0)} \pa_{\bar{z}}
    + v_{\bar{z}}^{(0)} \pa_z) v_z^{(0)})
    \end{array}
    \right\}.
    \label{kansei5}
\end{eqnarray}
The matching condition (\ref{4.1}) can be understood as
\begin{equation}
{\bf v}^{(0)} ({\bf x} = {\bf X}(t, \xi)) = \dot{\bf X} (t, \xi)
\label{kansei6}
\end{equation}
and
\begin{equation}
{\bf v}^{(1)} ({\bf x} = {\bf X}(t, \xi)) = 0
\label{kansei7}
\end{equation}

As an example we will take up the ciliated motion in the $D=2$ fluid.
Then ${\bf v}^{(0)}$ satisfying (\ref{kansei3.1}) and (\ref{kansei6})
reads as before
\begin{equation}
v_{\bar{z}}^{(0)} (z, \bar{z}) = v^{(-)}(z) + v^{(+)}(\bar{z})
+ (\bar{z}^{-1} - z) \overline{{v'}^{(-)}(z)}
\label{kansei8}
\end{equation}
where
\begin{eqnarray}
v^{(-)} (z) & = & \frac{1}{2} \left\{ \sum_{n<1} \dot{\alpha}_n z^n
- \sum_{n<1, m+n<2}
n \alpha_m \dot{\alpha}_n z^{m+n-1} + \sum_{n<1,m<n} n \alpha_m
\overline{\dot{\alpha}_n} z^{m-n+1}  \right. \nonumber \\
& & \left. + \sum_{n \geq 1, m>n} n \overline{\alpha_m}
\dot{\alpha}_n z^{-m+n+1} + \sum_{n<1, m+n>2} n
\overline{\alpha_m} \overline{\dot{\alpha}_n} z^{-m-n+3}\right\},    \\
v^{(+)}(\bar{z}) & = & \frac{1}{2}  \left\{ \sum_{n \geq 1}
\dot{\alpha}_n \bar{z}^{-n} - \sum_{n<1, m+n \geq 2} n \alpha_m
\dot{\alpha}_n \bar{z}^{-m-n+1} +
\sum_{n<1, m>n} n \alpha_m \overline{\dot{\alpha}_n}
\bar{z}^{-m+n-1}  \right.
\nonumber \\
& & \left. \sum_{n \geq 1, m \leq n} n \overline{\alpha_m}
\dot{\alpha}_n \bar{z}^{m-n-1} + \sum_{n<1, m+n \leq 2} n
\overline{\alpha_m} \overline{\dot{\alpha}_n}\bar{z}^{m+n-3}  \right\}.
\end{eqnarray}
Substitution of ${\bf v}^{(0)}$ into eq.(\ref{kansei3.2}),
we can find ${\bf v}^{(1)}$ satisfying eq.(\ref{kansei7})
as well as the boundary condition at spacial infinity
where ${\bf v}^{(1)}$ is at least finite.
This is carried out by adding properly the arbitrary
solution ${\bf v}^{(1)'}$
satisfying the homogeneous equation
$\Delta (\nabla \times {\bf v}^{(1)'})= 0$. We have found, however,
unwanted terms behaving
$z^2 \bar{z}^{-2}, \ \bar{z}^2 z^{-2}, \ \ln z $, and $\ln \bar{z}$
in the solution ${\bf v}^{(1)}$.
To eliminate these unwanted terms, we impose the following
restriction on $\alpha(s, t)$;
\begin{eqnarray}
 \lefteqn{ \overline{\ddot{\alpha}_k} + \sum_{n<0}
  n \{( \overline{\alpha_{-n+k+1}} \overline{\ddot{\alpha}_n}
 + \overline{\dot{\alpha}_{-n+k+1}} \overline{\dot{\alpha}_n} ) }  \nonumber \\
& &  -  ( \overline{\dot{\alpha}_{n+k-1}} \dot{\alpha}_n
+ \overline{\alpha_{n+k-1}} \ddot{\alpha}_n )
  -  ( \dot{\alpha}_{-n-k+3} \dot{\alpha}_n
  + \alpha_{-n-k+3} \ddot{\alpha}_n )  \}  \nonumber \\
 & & + \sum_{n \geq 1} n ( \dot{\alpha}_{n-k+1} \overline{\dot{\alpha}_n}
 + \alpha_{n-k+1} \overline{\ddot{\alpha}_n} ) \nonumber  \\
 & &  + \frac{2}{k} \sum_{n<1} (n-k-1)(n-k)
   \overline{\dot{\alpha}_n} \overline{\dot{\alpha}_{-n+k+1}} = 0,
   \hspace{1cm}  (k=-1, \  -2)\label{masako}
\end{eqnarray}
\begin{equation}
 \sum_{n<1} n(n+1) \dot{\alpha}_n \dot{\alpha}_{-n} = 0,
 \label{masako3}
\end{equation}
\begin{eqnarray}
& (k+1) ( \overline{\dot{\alpha}_2} \overline{\dot{\alpha}_{k+1}} +
 \dot{\alpha}_{-1} \overline{\dot{\alpha}_k} )= 0   &
 \hspace{1cm} (k<0)  \\
& \dot{\alpha}_1 \dot{\alpha}_k = 0 \hspace{1cm} (k<0) &
\end{eqnarray}
Now, we have the final solution of ${\bf v}^{(0)}+ {\bf v}^{(1)}$.
This gives the net swimming velocity $\tilde{v}_T^{(\rm cilia)}$
and the angular momentum gained $\tilde{v}_R^{(\rm cilia)}$ when
the \mos\ is swimming in the fluid with small but non-vanishing
Reynolds number;
\begin{eqnarray}
 \lefteqn{ \tilde{v}_T^{(\rm cilia)}  =  -\dot{\alpha}_0 (t) }
 \nonumber    \\
      &  & + \sum_{k \leq 1} k (\dot{\alpha}_k \alpha_{-k+1} -
      \overline{\dot{\alpha}_k} \alpha_{k-1} - \overline{\dot{\alpha}_k}
      \overline{\alpha_{-k+3}})
     - \sum_{k>1} k \dot{\alpha _k} \overline{\alpha _{k-1}}  \nonumber   \\
     & &  + \frac{\epsilon}{4} \left\{
     \left(
    \sum_{k>2} \frac{2(k-3)}{(k-1)(k-2)}  \overline{\dot{\alpha}_k}
    \overline{\dot{\alpha}_{-k+3}}  -
    \sum_{k>1} \frac{2}{k-1} \dot{\alpha}_k \dot{\alpha}_{-k+1} \right. \right.
     \nonumber  \\
   & &    \left. - \sum_{k<-1} \frac{2(3k+1)}{k(k-1)(k+1)}
    \dot{\alpha}_k \overline{\dot{\alpha}_{k+1}} )
    \right)      \nonumber  \\
   & & +  \sum_{k<-1} \frac{1}{k+1}
   \left(  \alpha_{-k+1} \ddot{\alpha}_k
   - \alpha_{k-1} \overline{\ddot{\alpha}_k}
   +  \overline{\alpha_{k+1}} \ddot{\alpha}_k -
   \overline{\alpha_{-k+3}}\overline{\ddot{\alpha}_k}  \right)
   \nonumber   \\
   & &  \left.  - (\alpha_{-2} \overline{\ddot{\alpha}_{-1}}
   + \overline{\alpha_4} \overline{\ddot{\alpha}_{-1}} ) \right\},
     \label{kansei11}
\end{eqnarray}
\begin{eqnarray}
 \lefteqn{ \tilde{v}_R^{(\rm cilia)}  =  -{\rm Im} \{ \dot{\alpha}_1(t)
    }   \nonumber      \\
    &  & - \left. \sum_{k \leq 1} k (\dot{\alpha}_k \alpha_{-k+2}
    - \overline{\dot{\alpha}_k} \alpha_{k} - \overline{\dot{\alpha}_k}
    \overline{\alpha_{-k+2}})      +  \sum_{k>1} k \dot{\alpha _k}
    \overline{\alpha _k} \right\} \nonumber   \\
    & &  + \frac{\epsilon}{4}{\rm Im}  \left\{
    \sum_{k>1} \frac{2}{k-1} (  \overline{\dot{\alpha}_{-k+2}}
    \overline{\dot{\alpha}_k} -
    \dot{\alpha}_{-k+2} \dot{\alpha}_k )
    -(\alpha_{-1} \overline{\ddot{\alpha}_{-1}} + \overline{\alpha_3}
   \overline{\ddot{\alpha}_{-1}} )  \right. \nonumber   \\
    & & \left. + \sum_{k<-1} \frac{1}{k+1}
    ( \alpha_{-k+2} \ddot{\alpha}_k
   - \alpha_k \overline{\ddot{\alpha}_k} + \overline{\alpha_k} \ddot{\alpha}_k
   -\overline{\alpha_{-k+2}} \overline{\dot{\alpha}_k}  )  \right\}
    \label{kansei12}
\end{eqnarray}
It is interesting to note that the selection rules (\ref{36.1}) and
(\ref{36.2}) are also valid in this
case of small but non-vanishing Reynolds number.

\section{Collective motion of microorganisms
and $N$-point correlation function}
Our previous study in this problem is summarized in the following.
Consider the situation where the vortices are created and annihilated,
so that the probability of having the vortex distribution (field)
$\omega^{\mu \nu}(x)$ is given by
\begin{equation}
   P[\omega^{\mu \nu}(x)] \sim \exp \left\{
   -\frac{1}{2 \pi i \alpha'} \int d^D x \sqrt{g(x)} \: \frac{1}{4}
   \omega_{\mu \nu}(x) \omega^{\mu \nu} (x)     \right\},
   \label{66}
\end{equation}
where $i \alpha'$ is the external parameter controlling the fluctuation
of the vortex distribution. [The $\alpha' \rightarrow 0$ limit corresponds
to the classical limit without the fluctuation.] In this situation
we should sum over all the possible configurations of the velocity
fields with Eq.(\ref{66}) as their probability. The probability of
having $N$ \mos s with their surfaces located at $X_{(1)}, X_{(2)},
\cdots, X_{(N)}$, and with their time derivatives
$\dot{X}_{(1)}, \dot{X}_{(2)}, \cdots,\dot{X}_{(N)}$, is given by the
following $N$-point correlation function (amplitude);
\begin{eqnarray}
   \lefteqn{ G_N (X_{(1)},\dot{X}_{(1)}; \, \cdots; \, X_{(N)},
   \dot{X}_{(N)}) }      \nonumber        \\
   &  & = \int {\cal D} P_{\mu}^{(i)} \, \exp \left\{
          i \sum_{i=1}^{N} \int dt \, \int d^{D-1}\xi_{(i)} \,
          P_{\mu}^{(i)}                \dot{X}^{\mu}  \right\}
          \nonumber     \\
   &  & \; \times  \: \tilde{G}_N ( X_{(1)}, P^{(1)}; \,
   \cdots \, ;X_{(N)}, P^{(N)} ),
      \label{69}
\end{eqnarray}
where
\begin{eqnarray}
   \lefteqn{  \tilde{G}_N ( X_{(1)}, P^{(1)}; \, \cdots
   \, ;X_{(N)}, P^{(N)} )}
   \nonumber     \\
   &   &  = \exp \left[ 2 \pi i \alpha' \times \frac{1}{2}
          \sum_{i, j} \int dt_{(i)} \, \int d^{D-1}\xi_{(i)} \,
           \int dt_{(j)} \, \int d^{D-1}\xi_{(j)}  \right.
           \nonumber    \\
    &   &  \; \times \left.  P_{\mu}^{(i)}(t_{(i)}; \xi_{(i)} )
          G_{\perp}^{\mu \nu} \left( X_{(i)}(\xi_{(i)}) - X_{(j)}(\xi_{(j)})
          \right)  P_{\mu}^{(j)}(t_{(j)}; \xi_{(j)} )  \right],
   \label{71}
\end{eqnarray}
with the Green's function of the transverse waves.
Then, we have obtained $N$-point correlation function for the
collective swimming of $N$ \mos s. It is quite similar to the
$N$-point function of strings for $D=2$ case and membranes for
$D=3$ case. It is also related to the string field theory,
since the incoming and outcoming strings are not point-like,
but the Reggeons.

Here we will discuss the use of $N$-point correlation function
(\ref{69}) in the collective swimming motion of \mos s. If $G_N$
represents the probability of having $N$ \mos s whose surfaces are
located at $X_{(i)}$ with velocity $\dot{X}_{(i)}\ (i=1, \cdots, N)$,
then it can be viewed as the probability distribution of $\dot{X}_N$
of the imaginary \mos\ $N$ located at spacial infinity under the given
data of $X_i$ and $\dot{X}_i (i=1, \cdots, N-1)$. Following the usual
strategy, the counterflow $-\dot{X}_N$ can be indentified to the
collective swimming motion of $N-1$ \mos s. Therefore, the averaged
collective swimming motion over the fluctuation distribution is given by
\begin{equation}
  -\langle \dot{X}_N \rangle = -\sum_{\dot{X}_N} \dot{X}_N
    G_N (X_1, \dot{X}_1; \cdots ; X_N, \dot{X}_N).
    \label{74}
\end{equation}
It is also an interesting problem to include the kinetic terms of the
\mos s
themselves, or to consider the stochastic behavior in the swimming motion
of the \mos s.

\section*{Summary}
In this paper, recaptulating our previous work on the swimming of \mos s,
we have given a few newly obtained results: we have obtained the closed
algebra and its central extension, which is a generalization of the
$W_{1 + \infty}$ algebra, controlling the swimming motion of \mos.
For the ciliated motion, we have extended our previous result to the
larger \mos s\ for which the Reynolds number is small but non-vanishing.
We have developed the perturbation theory with respect to the Reynolds number.

\section*{Acknowledgements}
This work is partly supported by Grant-in-Aid for Scientific Research from the
Ministry of Education, Science and Culture (No.06221229).

\end{document}